\documentclass[pra,twocolumn]{revtex4}
\usepackage{amsmath}
\usepackage{graphicx}

\begin{document}

\title{Stable two-dimensional solitons supported by radially inhomogeneous
self-focusing nonlinearity}
\author{Hidetsugu Sakaguchi$^{1}$ and Boris A. Malomed$^{2}$}
\address{$^{1}$Department of Applied Science for Electronics and
Materials, Interdisciplinary Graduate School of Engineering
Sciences, Kyushu
University, Kasuga, Fukuoka 816-8580, Japan \\
$^{*}$Corresponding author:sakaguchi@asem.kyushu-u.ac.jp\\
$^{2}$Department of Physical Electronics, School of Electrical
Engineering, Faculty of Engineering, Tel Aviv University, Tel Aviv
69978, Israel}

\begin{abstract}
We demonstrate that modulation of the local strength of the cubic
self-focusing (SF) nonlinearity in the two-dimensional (2D) geometry, in the
form of a circle with contrast $\Delta g$ of the SF coefficient relative to
the ambient medium with a weaker nonlinearity, stabilizes a family of
fundamental solitons against the critical collapse. The result is obtained
in an analytical form, using the variational approximation (VA) and
Vakhitov-Kolokolov (VK) stability criterion, and corroborated by numerical
computations. For the small contrast, the stability interval of the
soliton's norm scales as $\Delta N\sim \Delta g$ (the replacement of the
circle by an annulus leads to a reduction of the stability region by
perturbations breaking the axial symmetry). To further illustrate this
mechanism, we demonstrate, in an exact form, the stabilization of 1D
solitons against the critical collapse under the action of a locally
enhanced quintic SF nonlinearity. 
\end{abstract}

\maketitle

\noindent OCIS numbers: 190.6135, 190.3100, 190.3270,
190.4390\bigskip

The use of spatially modulated nonlinearities for supporting solitons in
optical waveguides and Bose-Einstein condensates (BECs) has drawn much
attention in past several years \cite{review,Barcelona}. Various patterns of
the local strength of the nonlinearity can be created in photonic crystals
\cite{1DKominisFirst}-\cite{Asia}, or with the help of inhomogeneous
distributions of resonant dopants in optical waveguides \cite{Kip}. Similar
structures can be induced in BEC by means of the Feshbach resonance
controlled by nonuniform external fields \cite{1Dwe}-\cite{2DSivan}. In the
latter case, an effective technique for creating necessary field patterns
may be provided by arrayed magnetic films \cite{magnets}.

The spatially modulated nonlinearity gives rise to effective \textit{%
pseudopotentials} \cite{pseudo} which help to create stable spatial solitons
\cite{review} in diverse one-dimensional (1D) settings with the cubic \cite%
{1DKominisFirst}-\cite{1DBarcelona}, \cite{1Dwe}-\cite{pseudo}, quadratic
\cite{Asia} and quintic \cite{Zeng} nonlinearities. Stabilizing
multidimensional solitons by means of similar techniques is a harder
problem. In the 2D geometry, smooth landscapes of the self-focusing (SF)\
cubic nonlinearity do not allow one to stabilize solitons against the
corresponding \textit{critical} \cite{Rub1}-\cite{Berge'} collapse \cite%
{2DSivan}. Nevertheless, the stabilization of 2D fundamental
solitons in sufficiently broad parametric regions is possible in
circles \cite{2Dwe,2Dcircles}
or annuli \cite{2Dwe}, as well as in stripe planforms \cite%
{2DMarek}, with \emph{sharp edges}. No results for stable 2D vortex
solitons, or 3D solitons of any kind, supported by inhomogeneous SF
nonlinearities have been reported thus far. On the other hand, the
stabilization of all kinds of bright solitons and solitary vortices
in any dimension $D$ is readily provided by the
\emph{self-defocusing} cubic nonlinearity, whose local strength is
made to grow at large distances ($r$) faster than $r^{D}$
\cite{SDF}.
A completely different mechanism for the stabilization of 2D solitons, based
on the use of localized gain \cite{gain}, is possible in dissipative optical
media.

The stabilization mechanism for 2D fundamental solitons has been
demonstrated in areas carrying the SF cubic nonlinearity and embedded into
self-defocusing or linear host media \cite{2Dwe}-\cite{2DMarek}. A
challenging question is whether the stabilization of solitons against the
critical collapse may be possible solely due of the contrast between
stronger SF in a given area, and a weaker self-attractive nonlinearity in
the ambient medium. The objective of this Letter is to demonstrate that such
a mechanism works for fundamental solitons trapped in circles and annuli.
Thus, we demonstrate the stabilization of the 2D solitons by the \emph{%
enhanced self-focusing}: while the well-known Townes solitons are unstable
against the collapse in the uniform SF medium \cite{Berge'}, the creation of
a circle or annulus with a larger SF coefficient, $g$, gives rise to a
family of \emph{stable} 2D solitons. In particular, in the limit of small
contrast $\Delta g$ between the circle or annulus and the host medium, the
stability interval for the solitons scales as $\Delta g$. To illustrate the
genericity of the mechanism, we also report exact results for the
stabilization of 1D fundamental solitons against the respective critical
collapse induced by the quintic nonlinearity.

The normalized form of the 2D nonlinear Schr\"{o}dinger (NLS) equation for
wave amplitude $\phi \left( x,y,z\right) $ in the medium with the modulated
nonlinearity is \cite{2Dwe}
\begin{equation}
i\phi _{z}=-(1/2)\nabla ^{2}\phi -g(r)|\phi |^{2}\phi ,  \label{GP}
\end{equation}%
where $z$ is the propagation distance, $x$ and $y$ are the transverse
coordinates, $r=\sqrt{x^{2}+y^{2}}$, and%
\begin{equation}
g(r)=\left\{
\begin{array}{c}
1,~\mathrm{at~}\rho <r<2, \\
1-\Delta g,~\mathrm{at}~~r<\rho ~\mathrm{and}~r>2,%
\end{array}%
\right.  \label{g}
\end{equation}%
$\rho $ being the inner radius of the annulus (the circle corresponds to $%
\rho =0$), while the outer radius is set to be $2$ by scaling. Inside the
annulus, the SF coefficient is normalized to be $1$, while $1-\Delta g$ is
the background value. In the application to BEC, the evolutional variable in
Eq. (\ref{GP}) is time $t$, instead of $z$.

Solutions to Eq. (\ref{GP}) in the form of axisymmetric localized modes with
propagation constant $-\mu >0$ are sought for as $\phi =u(r)\exp (-i\mu z)$,
with real $u(r)$ determined by equation
\begin{equation}
\mu u=-(1/2)\left( u^{\prime \prime }+r^{-1}u^{\prime }\right) -g(r)u^{3}.
\label{u}
\end{equation}%
It does not make sense to consider $\Delta g<0$, as in that case the annulus
is a repulsive structure, that cannot trap stable solitons. On the other
hand, $\Delta g>1$ implies that the host medium is self-defocusing, which,
repelling the wave field, helps to trap and stabilize modes in the SF
annulus \cite{2Dwe,2Dcircles}. In the case of $\Delta g\gg 1$, the field is
strongly confined to the annulus. Indeed, in a boundary layer adjacent to $%
r=2$, the solution to Eq. (\ref{u}) interpolating between the inner one (at $%
r<2$), which is characterized by the boundary value of the derivative, $%
u_{r=2}^{\prime }$, and the outer solution, which quickly decays at $r>2$,
is $u\approx \sqrt{-u_{r=2}^{\prime }}\left[ \left( \Delta g\right) ^{1/4}+%
\sqrt{\Delta g\left( -u_{r=2}^{\prime }\right) }\left( r-2\right) \right]
^{-1},$ with $u\left( r=2\right) \approx \left( \Delta g\right) ^{-1/4}\sqrt{%
-u_{r=2}^{\prime }}$. This means that, in the case of $\Delta g\rightarrow
\infty $, one should look for the inner solution satisfying the boundary
condition $u(r=2)=0$, with arbitrary $u_{r=2}^{\prime }$. The same pertains
to the boundary at $r=\rho $.

\begin{figure}[tbp]
\includegraphics[height=2.6cm]{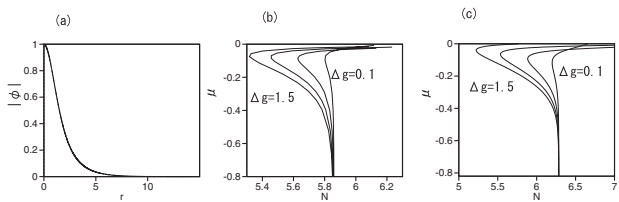}
\caption{(a) A stable localized solution for $\Delta g=0.5$, $\protect\rho %
=0 $ and $\protect\mu =-0.197,N=5.73$. (b) The propagation constant of
numerically found families of the trapped fundamental modes vs. the norm, at
fixed values of the nonlinearity contrast between the circle ($\protect\rho %
=0$) and surrounding background, $\Delta g=1.5,1,0.5$, and $0.1$. (c) The
same as in (b), but as obtained from the VA.}
\label{fig1}
\end{figure}

Numerical solutions of Eq. (\ref{u}) were found by means of the shooting
method. Figure \ref{fig1}(a) displays a typical profile of stable
fundamental (single-peak) modes. Families of the solutions are presented in
Fig. \ref{fig1}(b) by curves for the total power (norm), $N=2\pi
\int_{0}^{\infty }u^{2}(r)rdr$, versus $\mu $. All the families contain
portions that meet the Vakhitov-Kolokolov (VK) stability criterion, $d\mu
/dN<0$ \cite{Kol,Rub2,Berge'}. Direct simulations of the full 2D equation (%
\ref{GP}) (without imposing the axial symmetry) confirm that all the
solutions satisfying the VK criterion are stable, while those corresponding
to $d\mu /dN>0$ suffer the collapse or decay, as expected.

A natural tool for the consideration of the present setting is provided by
the variational approximation (VA) \cite{Anderson,2Dwe}. To this end, we
adopt the Gaussian ansatz, $u=A\exp \{-r^{2}/(2W^{2})\},$ with amplitude $A$%
, width $W$, and norm $N=\pi A^{2}W^{2}$. The substitution of this into the
underlying Lagrangian, $L=2\pi \int_{0}^{\infty }\left[ 2\mu u^{2}-\left(
u^{\prime }\right) ^{2}+g(r)u^{4}\right] rdr$, yields
\begin{equation}
L=2\mu N-\frac{N}{W^{2}}+\frac{N^{2}}{2\pi W^{2}}\left[1-\Delta g(1-E_2+E_1)\right],
\end{equation}
where $E_1=\exp(-8/W^2)$ and $E_2=\exp(-2\rho^2/W^2)$.
Then, $W$ and $\mu $ are determined by the variational equations,
$\partial L/\partial W=\partial L/\partial N=0$.
Figure \ref{fig1}(c) shows that the VA produces curves $\mu (N)$
which are very close to their numerical counterparts. Further, a
region in which the VA predicts localized modes that are stable as
per the VK criterion (hence they are stable indeed, as confirmed by
the aforementioned numerical results) can be found from the
variational equation $\partial L/\partial W=0$:
\begin{equation}
\left( 1+e^{-2}\Delta g\right) ^{-1}<N/(2\pi )<1  \label{int}
\end{equation}%
($N=2\pi $ is the prediction of the VA for the norm of the Townes
soliton)
\cite{Anderson}). Note that the smallest value of $N$ in Eq.
(\ref{int}) corresponds to width $W=2$, which coincides with the
radius of the circle, the stable branches of the $\mu (N)$ curves in
Fig. \ref{fig1}(c) corresponding to $W\leq 2$, i.e., the stable
modes are trapped inside the circle. Further, it is seen from Eq.
(\ref{int}) that, for the vanishing contrast between the circle and
background, $\Delta g\rightarrow 0$, the width of the stability
interval shrinks as $\Delta N\approx 2\pi e^{-2}\Delta g\approx
0.85\Delta g$. The analysis of numerical solutions yields a very
close result for $\Delta N$. Thus we conclude that there exists the
\emph{stable} subfamily of the 2D solitons in the circle, as long as
the SF profile is not completely flat.

\begin{figure}[tbp]
\includegraphics[height=2.6cm]{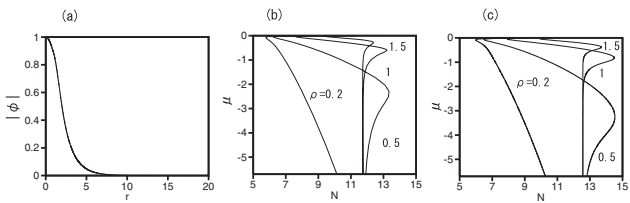}
\caption{(a) The profile of a fundamental soliton trapped in the annulus
with $\protect\rho =1$, $\Delta g=0.5$ and $\protect\mu =-0.216$, $N=9.82$.
(b) The propagation constant of numerically found families of the trapped
modes vs. $N$ in annuli with nonlinearity contrast $\Delta g=0.5$ and
different inner radii $\protect\rho $. (c) The same as in (b), but as
produced by the VA.}
\label{fig2}
\end{figure}

Results for the annuli are shown in Fig. 2. The fundamental soliton still
features a maximum at $r=0$ in Fig. \ref{fig2}(a), in spite of the
relatively large radius and depth of the inner hole in that case, $\rho =1$
and $\Delta g=0.5$. For a small inner radius $\rho $, the VA yields
solutions in the range of $\left( 1+e^{-2}\Delta g\right) ^{-1}<N/(2\pi
)<\left( 1-(1+e^{-2})\Delta g\right) ^{-1}$, cf. Eq. (\ref{int}). However,
simulations demonstrate that the fundamental modes trapped in the annulus
are stable only in a narrow interval near the left edge of this region---for
instance, at $5.78<N<5.92$ for $\rho =0.2$, $\Delta g=0.5$. At larger $N$,
the modes are destabilized by perturbations which break the axial symmetry,
driving the soliton's peak off the center and, eventually, causing the
collapse. This instability (similar to that found for the annulus embedded
into the linear medium \cite{2Dwe}) is not comprised by the VK criterion,
which pertains solely to axisymmetric perturbations.

We have also checked the dynamics of modes with embedded vorticity in this
2D setting, concluding that, as well as in other cases when the transverse
or longitudinal modulation of the SF\ cubic nonlinearity is employed \cite%
{2Dwe}, \cite{2Dcircles}, \cite{Caputo}-\cite{Centurion}, vortex solitons
cannot be stabilized against splitting by azimuthal perturbations (on the
other hand, solitary vortices supported by the transverse modulation of the
self-defocusing nonlinearity growing at $r\rightarrow \infty $ can be easily
made stable \cite{SDF}).

The 1D counterpart of the critical collapse occurs in the NLS equation with
the quintic term, $i\phi _{z}=-(1/2)\phi _{xx}-|\phi |^{4}\phi $ \cite%
{Abd,Alf,Zeng}. In optics, this model can be realized experimentally in
colloids \cite{colloid}.%
The quintic equation gives rise to the 1D version of the Townes
soliton,
$\phi =\left( -3\mu \right) ^{1/4}e^{-i\mu z}\sqrt{\mathrm{%
sech}\left( \sqrt{-8\mu }x\right) }$, with the constant total power:
$N_{0}\equiv \sqrt{3/8}\pi $ for any $\mu <0$. The simplest 1D
counterpart of the 2D setting considered above is provided by the
spatially modulated coefficient in front of the quintic term:
$g(x)=1+\Delta g\cdot \delta (x)$, with $\Delta g>0$ [for a similar
radial modulation of the cubic SF nonlinearity in the 2D model,
$g(r)=g_{0}\delta \left( r-r_{0}\right) $, exact solutions for
trapped modes can be found, but they are unstable against azimuthal
perturbations \cite{2Dwe}]. Then, the 1D Townes solitons are
replaced by exact
solutions $\phi =\left( -3\mu \right) ^{1/4}e^{-i\mu z}\sqrt{\mathrm{sech}%
\left( \sqrt{-8\mu }\left( |x|+\xi \right) \right) }$, with $\xi $ defined
by relation $\sinh \left( 2\sqrt{-8\mu }\xi \right) =3\Delta g\sqrt{-2\mu }$%
, and total power $N_{\mathrm{1D}}=\sqrt{6}\tan ^{-1}\left\{ \sqrt{\sqrt{%
-18\left( \Delta g\right) ^{2}\mu +1}+3\Delta g\sqrt{-2\mu }}\right\} $
taking values $\sqrt{3/8}\pi <N_{\mathrm{1D}}<\sqrt{3/2}\pi $. The entire
family satisfies the VK stability condition, $dN/d\mu <0,$ and is indeed
stable.

In conclusion, it is demonstrated that the axisymmetric modulation of the
strength of the SF cubic nonlinearity with sharp edges stabilizes a family
of 2D solitons against the critical collapse. For small contrast $\Delta g$, 
the width of the stability interval of the soliton's norm scales as $\Delta N\sim \Delta g$. For the annulus, the stability interval is strongly reduced by azimuthal perturbations. 

\end{document}